\documentclass[twoside]{article}

\usepackage{amssymb}
\usepackage{epsfig}

\newcommand{\nc}{\newcommand}
\nc{\be}{\begin{equation}}
\nc{\ee}{\end{equation}}
\nc{\bea}{\begin{eqnarray}}
\nc{\eea}{\end{eqnarray}}

\nc{\eqn}[1]{{(\ref{#1})}}
\nc{\cA}{{\cal A}}
\nc{\cB}{{\cal B}}
\nc{\cC}{{\cal C}}
\nc{\cD}{{\cal D}}
\nc{\cE}{{\cal E}}
\nc{\cF}{{\cal F}}
\nc{\cG}{{\cal G}}
\nc{\cH}{{\cal H}}
\nc{\cI}{{\cal I}}
\nc{\cJ}{{\cal J}}
\nc{\cK}{{\cal K}}
\nc{\cL}{{\cal L}}
\nc{\cM}{{\cal M}}
\nc{\cN}{{\cal N}}
\nc{\cO}{{\cal O}}
\nc{\cP}{{\cal P}}
\nc{\cQ}{{\cal Q}}
\nc{\cR}{{\cal R}}
\nc{\cS}{{\cal S}}
\nc{\cT}{{\cal T}}
\nc{\cU}{{\cal U}}
\nc{\cV}{{\cal V}}
\nc{\cW}{{\cal W}}
\nc{\cX}{{\cal X}}
\nc{\cY}{{\cal Y}}
\nc{\cZ}{{\cal Z}}
\newcommand{\AmS}{{\protect\the\textfont2
  A\kern-.1667em\lower.5ex\hbox{M}\kern-.125emS}}

\title{Finite size scaling analysis with linked cluster expansions
       \thanks{Talk presented by H.~Meyer-Ortmanns}}

\author{H. ~Meyer-Ortmanns \thanks{e-mail address:
        ortmanns@thphys.uni-heidelberg.de} 
        and T. ~Reisz \thanks{Heisenberg fellow, e-mail address: 
         t.reisz@thphys.uni-heidelberg.de} \\
        Institute of 
        Theoretical Physics \\ 
        University of Heidelberg \\ 
        Philosophenweg 16 \\
        D-69120 Heidelberg, Germany}
       
\begin{document}

\maketitle

\begin{abstract}
Linked cluster expansions are generalized from an infinite to a finite
volume on a $d$-dimensional hypercubic lattice. They are performed to
20th order in the expansion parameter to investigate the phase
structure of scalar $O(N)$ models for the cases of $N=1$ and $N=4$ in
3 dimensions. In particular we propose a new criterion to distinguish
first from second order transitions via the volume dependence of
response functions for couplings close to but not at the critical
value. The criterion is applicable to Monte Carlo simulations as
well. Here it is used to localize the tricritical line in a $\Phi^4 +
\Phi^6$ theory. We indicate further applications to the electroweak
transition.
\end{abstract}

\section{LINKED CLUSTER EXPANSIONS IN THE INFINITE VOLUME}

Convergent expansions such as linked cluster, hopping parameter or
high temperature expansions provide an analytic alternative to Monte
Carlo simulations. Originally they have been developed in the infinite
volume, meanwhile they have been extended to finite volumes as well
\cite{hilde}. In contrast to generic perturbation theory, hopping
parameter expansions (HPEs) are convergent expansions about completely
disordered lattice systems. The expansion parameter $\kappa$ is the
coefficient of the (pair) interaction term. 
Since we calculate free energies and connected
correlations in the hopping parameter expansion, we generate
linked cluster expansions (LCEs).

LCEs have been
generalized to scalar field theories at {\it finite} temperature by
\cite{reisz}. Compared to the zero temperature results,
the finite temperature induces a tiny shift in the critical
temperature (hopping parameter). Thus one has to face a similar
problem as the improvement of a bad signal to noise ratio in Monte
Carlo simulations. Here the signal corresponds to the finite
temperature effect and the noise to the "background" of temperature
zero graphs. The price to describe the critical region, in particular
critical exponents, with an accurracy of $\approx 1$ \% is to go to the
18th  order in
the expansion in $\kappa$. This accurracy has been achieved in
\cite{reisz}. Thus the question arises, why we are still interested in
a generalization of LCEs to a finite volume. The reason is twofold.

$\bullet$ The first one is to identify 1st order transitions from series
expansions in the high temperature phase. For example a measurement of
a vanishing critical exponent $\nu \cdot \eta$ is compatible with a
Gaussian fixed point, but not conclusive for the onset of a first
order transition region at a tricritical point. (The latter
interpretation actually applies to our measurements.)

$\bullet$  The second reason is to distinguish second order transitions
associated with different universality classes. Ideally critical
exponents should show a clear gap between plateaus of different
universality classes. Practically the gap is smeared out because of
the truncation of the series expansions at finite order. The
truncation has a similar rounding effect on this gap as a finite volume
on thermodynamic singularities.

\section{GENERALIZATION OF HPEs TO A FINITE VOLUME}

\subsection{Embedding factors on the lattice}

In HPEs the action is split into a sum of ultralocal parts
$\stackrel{\circ}{S}$ 
and a next neighbour part $S_{nn}$ with
next neighbour couplings $\propto \kappa$. 
A Taylor expansion in $\kappa$ of the partition
function $Z$ about the ultralocal contribution to $Z$ finally leads 
to graphical
expansions of n-point susceptibilities $\chi_n$
\be \label{exa.chi}
   \chi_{n}(\kappa) =   \sum_{\mu\geq 0} a_\mu^{(n)} \kappa^\mu .
\ee
The coefficients $a_\mu^{(n)}$ are a sum of graphs each of which
consists of a product of the
inverse symmetry factor, an internal symmetry factor, a
lattice embedding factor, and a polynomial of vertex contributions 
depending on the couplings involved in $\stackrel{\circ}{S}$. It is only the
embedding factor that depends on the topology of the particular
lattice, that will change in passing from an infinite to a finite
volume.

%

\subsection{Shift in the critical coupling}


From our series expansions we measure a {\it shift in $\kappa_c$} from a fit
according to
\be \label{abc}
  \ln(| \kappa_c(L) - \kappa_c(\infty) |) \simeq   \ln{c} - y_T \ln{L}.
\ee
Here $c$ is a constant, $y_T=1/\nu$ for 2nd order transitions and
$y_T  \propto d$ for 1st order transitions. The critical couplings 
$\kappa_c(L)$ and $\kappa_c(\infty)$ are
determined as the radius of convergence of the series expansions
in a finite or infinite volume, respectively.
Since an extrapolation of the scaling behaviour from the height and
width of the critical region is not feasible within the series
expansions, we have proposed the following {\it monotony criterion}.

\subsection{The monotony criterion}

Consider two volumes $V_1$ and $V_2$
such that $V_1< V_2 \le \infty$, $0<\alpha<1$, $\sigma$ stands for a generic
coupling, and $t=(\kappa-\kappa_c)/\kappa_c$ for the reduced
"temperature".
 We define

\be \label{fss.rml}
   r(V_1,V_2) := 1 - \frac{ \chi_2(\kappa=\alpha \cdot \kappa_c,V_1)}
    {\chi_2(\kappa=\alpha \cdot \kappa_c,V_2)}  ,
\ee

The monotony criterion says that

\[ \label{cde}
    r_{V_1,V_2} \; \left\{ 
    \begin{array}{r@{\; ,\;} l }
    > 0 & \mbox{2nd order} \\
    < 0 & \mbox{1st order }\\
    = 0 & \mbox{tricritical point for 
         $\partial r/ \partial \sigma \neq 0$.}
    \end{array} \right. 
\]

The different behaviour of $r_{V_1,V_2}$ comes from the {\it singular}
part of $\chi_2$.
The underlying observation is a different monotony behaviour of
response functions $\chi$ for 1st and 2nd order transitions that is
seen for $\chi$s with nonanalytic behaviour in the infinite volume
limit. The $\chi$s are increasing in volume in a certain neighbourhood
of $T_c$ for 2nd order transitions  and decreasing for 1st order
transitions. For a precise specification of the scaling region,
i.e. the bounds that $V_1$, $V_2$,$|t|$ have to satisfy, we
refer to \cite{hilde}. The difference reflects the $\delta$-function
and power law singularities for 1st or 2nd order transitions,respectively,
predicted at $T_C$ in the thermodynamic limit.Thus the actual 
behaviour of $\chi_2$ we
have found in applications to scalar $O(N)$ models (cf. section 3)
is neither a pecularity of the model nor an artifact of the series
expansions.


$\bullet$  It should be noticed that the criterion is not restricted to series
expansions, but also applicable to Monte Carlo calculations (since
both involved volumes may be finite) if the regular contribution to
$\chi$ is really negligible.

$\bullet$  It is neither restricted to order parameter susceptibilities as the
notation suggests, but similarly applies to other response functions
like the specific heat if they diverge in the thermodynamic limit.

\subsection{The effective potential for $V<\infty$}

The effective potential is defined via the effective action $\Gamma$
evaluated for a constant background field 
$M$ ($L^d \cdot V_{eff}:= -\Gamma(M)$). It can be
expressed in terms of the 1PI-susceptibilities according to 
\[
  V_{eff}(M) =
  \frac{1-4D\kappa \chi_2^{1PI}}{\chi_2^{1PI}}
   \frac{M^2}{2}
   - \frac{\chi_4^{1PI}}{(\chi_2^{1PI})^4}
   \frac{(M^2)^2}{4!}
\]
\[
  -
  \frac{1}{(\chi_2^{1PI})^6}
  \; \left( \chi_6^{1PI} - \frac{10 (\chi_4^{1PI})^2}{\chi_2^{1PI}}
  \right) \frac{(M^2)^3}{6!}
  + O(M^8).
\]
The $\chi_{2n}^{1PI}$s are obtained from the $\chi_{2n}$s if the
graphical expansion of the $\chi_{2n}$s is restricted to 1PI graphs.
The (non)convex shape of the effective potential in the symmetric
phase indicates the order of the transition. Its evaluation in a
{\it finite} volume leads to an alternative determination of $\kappa_c(L)$
via the coexistence of different minima of the free energy 
or a vanishing of the coefficient of the quadratic term. 

\section{RESULTS FOR A $3d$ $\Phi^4 + \Phi^6$ THEORY}

We apply the finite size scaling analysis with HPEs
to a scalar theory with $\Phi^4$ and $\Phi^6$-terms and
$O(N)$ symmetry in 3 dimensions, $N$ is chosen as $1$ and
$4$. After a suitable reparametrization the ultralocal part of the
action reads
\be \label{exa.action}
   \stackrel{\circ}{S}(\Phi,\lambda,\sigma)
    \; = \; \Phi^2 + \lambda (\Phi^2-1)^2
   + \sigma (\Phi^2-1)^3
\ee
The hopping parameter term is given as

\be \label{exa.part}
   S_{nn} = -2\kappa
    \sum_{x,y\;{\rm nn}} \Phi(x) \cdot
      \Phi(y).
\ee

For a suitable choice of couplings the model has a tricritical line
separating regions of first and second order transitions. Thus it
serves as an appropriate toy model for testing our methods.

A fit of the shift in $\kappa_c$ according to Eq.\ref{abc} leads
to the values of $y_T$ listed in the Table.\vskip10pt 
 


\label{tab:kappal}
\begin{tabular}{rrrr}   
\hline                
                  \multicolumn{1}{r}{$N$}      
                  & \multicolumn{1}{r}{$y_T$}
                 & \multicolumn{1}{r}{$$} 
                 & \multicolumn{1}{r}{order}\\ 
\hline
$1$    & $ 6.068(43)$ & $\approx 2d$ & 1st \\
$4$    & $ 5.55(59)$ & $\approx 2d$ &  1st \\
$1$         & $2.656(48)$ & $\approx 2/\nu$ & 2nd \\
$4$               & not conclusive &\qquad &2nd  \\
\hline
\end{tabular}

\vskip10pt
The value for
$y_T$ for the first order transitions confirms results of
\cite{borgs1} for $N=1$ and extends results of \cite{borgs2} for
$N=4$. It disagrees with the naive expectation of a scaling with $d$
rather than $2d$.
The figure shows results for the monotony 

\begin{figure}[htb]
\setlength{\unitlength}{0.8cm}
\begin{picture}(5.0,7.0)
\epsfig{file=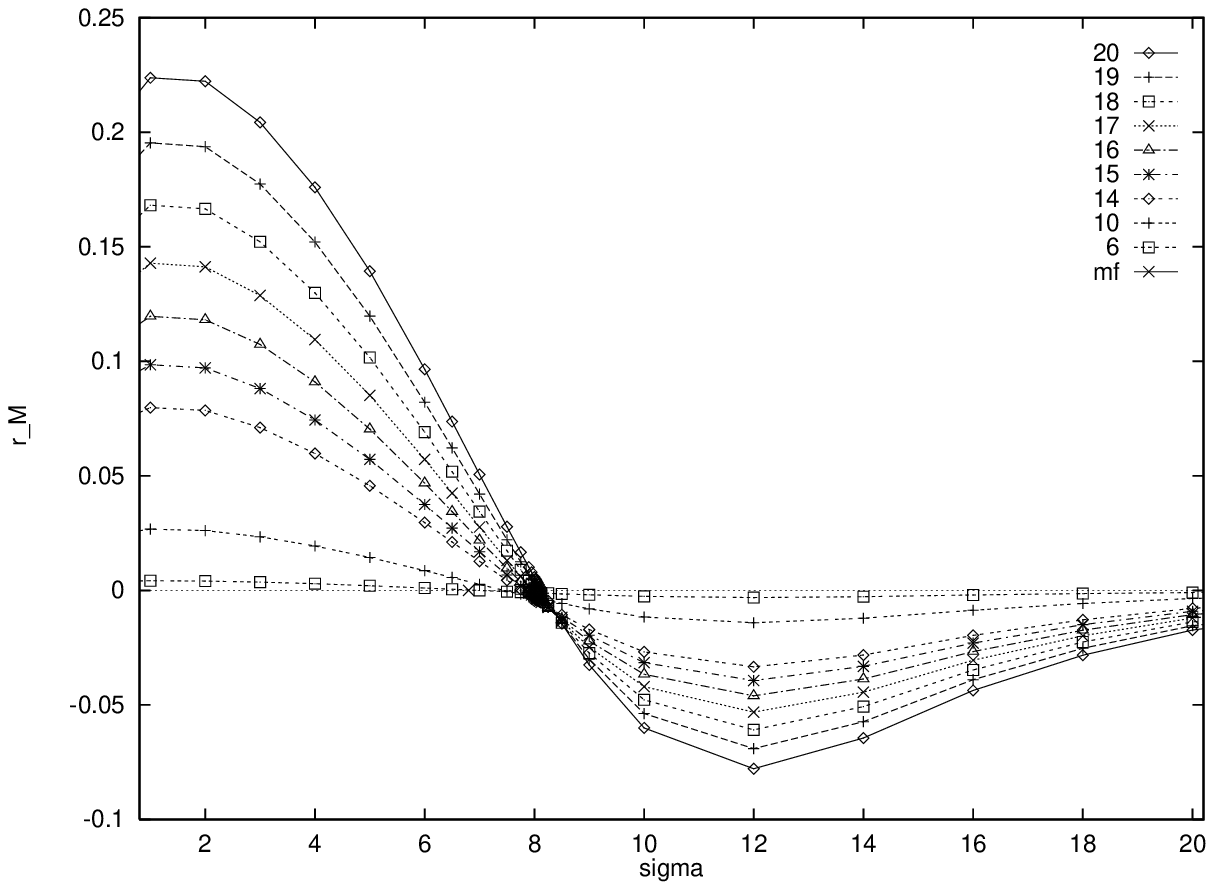,height=6.0cm,width=7.0cm,clip=}
\end{picture}
\end{figure}


criterion applied to the special case of $V_1=4^3$, $V_2=\infty^3$,
$\alpha=0.98$, $N=4$, $\lambda=\sigma/2$. The additional index $M$
refers to the order of truncation. 
A similar figure is obtained for $r(V_1,V_2)$ 
as function of $\sigma$
for fixed order of the truncation and varying volumes. Thus 
the intersection of the curves with the $\sigma$-axis, which
localizes the tricritical point, weakly depends on the order of the
truncation and the choice of $V_1$. The ultimate determination
of the tricritical couplings $\sigma_{t}$ and $\lambda_{t}$ needs
an extrapolation in both parameters. Respecting the mutual
relation between the truncation effect and the finite volume dependence
of the intersection, the final estimate for $\sigma_{t}$ leads to
$\sigma_t = 9.454(49)$ .
For further details on the involved extrapolations we refer to
\cite{hilde}. This result improves the infinite volume estimate of $8 \le
\sigma_{t} \le 10$ by two orders of magnitude.
It should be noticed that the volume independence of $\chi_2$ along the
tricritical line supports the validity of a mean field analysis of tricritical
exponents even in three dimensions as one might have expected from
the Ginzburg criterion. 
The localization of $\sigma_{t}$ from the change in the shape
of the effective potential leads to an upper
bound on $\sigma_{t}$, we find
  $9.75 \leq \sigma_t \leq 10.0$,
since the involved systematic errors are less well under control.
\vskip10pt
An extension to  
{\it effective} scalar models for an underlying
Salam-Weinberg theory with HPEs in a finite volume 
will particularly well work in the range of
large Higgs masses. It complements the range of Higgs masses
which has been available in recent Monte Carlo simulations and
allows a determination of the {\it critical} Higgs mass
above which the electroweak transition turns into a smooth crossover 
phenomenon \cite{nucu}.

%

\end{document}